\def\imat  {{\rm i}}
\documentstyle[twocolumn,floats,epsf,ifthen,aps,prb]{revtex} 

\begin{document} 

 

\title{Analysis of the radio-frequency single-electron transistor 
with large quality factor} 

\author{Valentin O. Turin 
and Alexander N. Korotkov}  
\address{ 
Department of Electrical Engineering, University of California, 
Riverside, CA 92521-0204. 
} 
\date{\today} 
 
\maketitle 
 
\begin{abstract} 
We have analyzed the response and noise-limited sensitivity 
of the radio-frequency single-electron transistor (RF-SET), extending the 
previously developed theory to the case of arbitrary large 
quality factor $Q$ of the RF-SET tank circuit. It is shown that while 
the RF-SET response reaches the maximum at $Q$ roughly corresponding 
to the impedance matching condition, the RF-SET sensitivity monotonically
worsens with the increase of $Q$. 
Also, we propose a novel operation mode of the RF-SET, in which an 
overtone of the incident rf wave is in resonance with the tank 
circuit. 
\end{abstract} 
 
\narrowtext 
 
\vspace{0.6cm}

\begin{figure}
\centerline{ 
\epsfxsize=3.5in
\hspace{-0.3cm}
\epsfbox{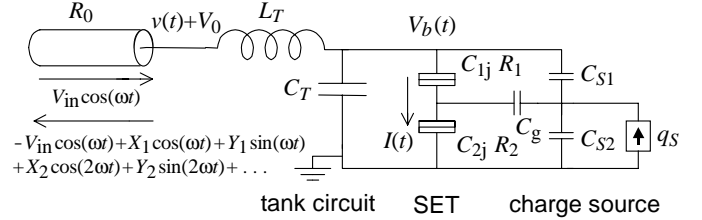} 
}
\caption{Schematic of the RF-SET. 
} 
\label{fig1}\end{figure}

        The problem of relatively small bandwidth of the 
conventional single-electron transistor\cite{Av-Likh,Fulton} (SET) 
due to its large output resistance, has been solved for many applications
by the invention\cite{Schoelkopf} of the radio-frequency SET (RF-SET),
which in many instances has already replaced the traditional SET setup. 
The principle of the RF-SET operation is somewhat similar to the operation
of the radio-frequency superconducting quantum interference device
\cite{Clarke} (RF-SQUID) and is based on the microwave reflection
\cite{Schoelkopf,Aassime,Aassime-2,Stevenson} from a tank circuit 
containing the SET (Fig.\ \ref{fig1}), which affects the quality factor 
($Q$-factor) of the tank; another possibility is to use 
the transmitted wave.\cite{Fujisawa,Cheong} 
The wide bandwidth of the RF-SET is due to the signal 
propagation by the microwave instead of charging the output wire,
while the tank circuit provides a better match between 
the cable wave impedance $R_0 = 50 \, \Omega$ and much larger 
SET resistance ($\sim 10^5\, \Omega$).

        The RF-SET bandwidth over 100 MHz has been demonstrated 
\cite{Schoelkopf} using the microwave carrier frequency $\omega /2\pi
=1.7$ GHz and relatively low $Q$-factor $Q\simeq 6$. However, in the 
present-day experiments the bandwidth is typically about 10 MHz
because of lower carrier frequency (to reduce amplifier noise)
and higher $Q$-factor (as an example, the bandwidth of 7 MHz for
$\omega /2\pi =332$ MHz and $Q>20$ has been reported in Ref.\
\cite{Aassime-2}). 

        Since the SET sensitivity is limited by the $1/f$ noise
only at frequencies below few kHz, the RF-SET typically operates
in the frequency range of shot-noise limited sensitivity of
the SET.\cite{Kor-92,Kor-94} The RF-SET charge sensitivity of 
$3.2\times 10^{-6} e/\sqrt{\mbox{Hz}}$ ($4.8\hbar$ in energy units) 
at  2 MHz has been reported in Ref. \cite{Aassime-2}. Even though
this value still contains comparable contributions from the SET and 
amplifier noises, the pure shot-noise-limited sensitivity seems to become
achievable pretty soon. 

     In spite of significant experimental RF-SET activity, we are 
aware of only few theoretical papers on the RF-SETs. 
The basic theory of the shot-noise-limited sensitivity of the RF-SET 
has been developed in Ref.\ \cite{Kor-RF}. A similar theory has been applied
to the analysis of the sensitivity of the RF-SET-based micromechanical 
displacement
detector.\cite{Blencowe,Zhang} Some theoretical analysis of 
the transmission-type RF-SET can be found in Ref.\ \cite{Cheong}. 

        In this letter we extend the theory of Ref.\ \cite{Kor-RF} 
to the case of arbitrary large $Q$-factor of the tank circuit, removing
the assumption of $Q$ being much smaller than the impedance-matching value. 
(While this condition was satisfied in the first experiment,\cite{Schoelkopf}
it is strongly violated in the present-day experiments.) 
We calculate the response and sensitivity of the normal-metal RF-SET and
find the optimal values numerically. Besides the usual case of the 
carrier wave in resonance with the tank circuit, we also consider the regime  
of a resonant overtone and find a comparable RF-SET performance 
in this case.

    We consider a SET (Fig.\ \ref{fig1}) consisting of two tunnel junctions
with capacitances $C_{1j}$ and $C_{2j}$ and resistances $R_1$ and $R_2$. 
The measured charge source $q_S$ has the capacitance $C_S=C_{S1}+C_{S2}$ 
and is coupled to the SET via capacitance $C_g$. Assuming constant $q_S$
(neglecting backaction), the SET can be reduced to 
the effective double-junction SET with capacitances 
$C_1= C_{1j}+C_gC_{S1}/(C_g+C_S)$, $C_2= C_{2j}+C_gC_{S2}/(C_g+C_S)$ and 
background charge $q_0=q_{00}+q_SC_g/(C_g+C_S)$, 
where $q_{00}$ is the initial contribution.
We will calculate the RF-SET response and sensitivity in respect to $q_0$,
while the corresponding quantities in respect to the measured charge $q_S$ 
differ by the factor $C_g/(C_g+C_S)$. 

    The current $I(t)$ through the SET affects the quality 
factor of the tank circuit consisting of the capacitance $C_T$ and
inductance $L_T$. In the linear approximation the SET can be  
replaced by an effective resistance $R_d$, and the total (``loaded'')
quality factor $Q_L=(1/Q+1/Q_{SET})^{-1}$ has contributions from
the ``unloaded'' $Q$-factor $Q=\sqrt{L_T/C_T}/R_0$ and damping by 
the SET: $Q_{SET}=R_d/\sqrt{L_T/C_T}$. For the incoming voltage wave 
${\hat V}_{in}\exp (\imat \omega t)$, the reflected wave 
$\alpha {\hat V}_{in}\exp (\imat \omega t)$ depends on the reflection
coefficient $\alpha =(Z-R_0)/(Z+R_0)$, where $Z=\imat \omega L_T
+(\imat \omega C_T+1/R_d)^{-1}$; close to the resonance, $\omega \approx
\omega_0=(L_TC_T)^{-1/2}$, it can be approximated as 
$Z\approx L_T/C_TR_d +2\imat (L_T/C_T)^{1/2}\Delta\omega /\omega_0$, 
$\Delta \omega \equiv \omega -\omega_0$.
Since an increment of the measured charge $q_S$ leads to an increment 
of $R_d$, the RF-SET response is proportional to $d\alpha /dR_d$. 
However, the amplitude ${\hat V_b}={\hat V}_{in}(Z+R_0)(\imat\omega C_T
+1/R_d)/2$ of the SET bias voltage oscillations should be determined by 
the Coulomb blockade threshold; so a more representative quantity
is 
        \begin{equation}
\frac{d\alpha}{dR_d} \frac{\hat{V}_{in}}{\hat{V}_b}
\approx \frac{-\imat R_0}{R_d^2}\, 
\frac{Q}{1+Q^2 R_0/R_d}\,
\frac{1}{1+2\imat Q_L\Delta\omega /\omega_0}. 
        \label{resp-lin}\end{equation}
This equation shows that the RF-SET response reaches the maximum 
at $Q=(R_d/R_0)^{1/2}$, which is the case of matched 
impedances at resonance, $Z\approx R_0$, and corresponds to the 
condition $Q=Q_{SET}=2Q_L$. 

        The linear analysis can only be used as an 
estimate because of the significant nonlinearity of the SET $I-V$ 
dependence. For the full analysis we use the differential 
equation \cite{Kor-RF} 
$ \ddot{v}/\omega_0^2 +\dot{v}/Q\omega_0 + v = 
2(1-\omega^2 /\omega_0^2)V_{in}\cos{\omega t}
-R_0 [I(t)-\langle I\rangle ]$, where $v(t)=V_{in}\cos \omega t+V_{out}(t)$ 
is the voltage at the end of the cable with subtracted dc component $V_0$ 
(see Fig.\ \ref{fig1}; do not use complex representation any more).  
The SET current $I(t)$ and its average
value $\langle I\rangle$ are found self-consistently 
from the SET bias voltage 
$V_b(t) = V_0 + v (t) + [2V_{in}\omega \sin{\omega t} +
\dot{v}(t)]\, Q/\omega_0$ using the ``orthodox'' model\cite{Av-Likh} 
and assuming continuous SET current ($\omega \ll I/e$). 

    In the steady state the reflected wave can be represented as
$V_{out}(t) = -V_{in} \cos{\omega t} +\sum_{n=1}^\infty
[ X_n \cos{n\omega t} +Y_n \sin{n\omega t} ]$ and the coefficients $X_n$ 
and $Y_n$ can be calculated as 
        \begin{eqnarray}
&& X_n=\{ R_0 Q  [n\tilde{\omega}\,a_n -Q (1-n^2\tilde{\omega}^2) b_n] 
+ 2 Q^2 (1-\tilde{\omega} ^2)^2 
\nonumber \\
&& \hspace{1.0cm} 
\times  V_{in}\delta_{1n} \} 
/ [n^2\tilde{\omega}^2 + Q^2(1 -n^2\tilde{\omega}^2)^2] , 
\label{Xn}
        \\ 
&& 
 Y_n=\{ - R_0 Q [n\tilde{\omega} \, b_n + Q (1-n^2\tilde{\omega}^2) a_n]
+  2 Q  \tilde{\omega} (1-\tilde{\omega}^2) 
\nonumber \\ 
&& \hspace{1cm} 
\times V_{in}\delta_{1n} \} 
/ [n^2\tilde{\omega}^2 + Q^2(1 -n^2\tilde{\omega}^2)^2] , 
        \label{Yn}
        \\ 
&& a_n =2 \langle I(t)\sin{n\omega t} \rangle , \,\,\,
b_n =2 \langle I(t)\cos{n\omega t} \rangle  , 
        \label{anbn}
        \end{eqnarray} 
where $\tilde{\omega}\equiv \omega /\omega_0$, 
$\delta_{1n}$ is the Kronecker symbol, 
and averaging is over the oscillation period, 
while $I(t)$ is determined by the SET voltage 
$V_b(t) = V_0 + 2Q\tilde\omega V_{in} \sin{\omega t}
+\sum_{n=1}^\infty [ (X_n+ Qn\tilde\omega Y_n) 
\cos{n\omega t} 
+ (Y_n- Qn\tilde\omega X_n) \sin{n\omega t} ]$.
Notice that the linear approximation corresponds
to neglecting the contribution of overtones ($n\geq 2$);
then $R_d =\pi A / [\int_0^{2\pi} I(V_0 + A\sin{x}) \sin{x}\, dx]$,
where $A$ is the amplitude of $V_b$ oscillations,
$V_b(t)=V_0 +A\sin (\omega t+\phi)$,
while there is no effective reactance contribution due to SET current. 
We used the self-consistent linear approximation 
as a starting point for the iterative solution of Eqs.\ 
(\ref{Xn})--(\ref{anbn}). 

    The RF-SET response in respect to monitoring the quadrature 
component $X_n$ can be defined as a derivative $dX_n/dq_0$ (similarly, 
$dY_n/dq_0$ for $Y_n$ monitoring). Other experimentally relevant
definitions are for monitoring 
the optimized phase-shifted combination $X_n\cos \varphi +Y_n\sin \varphi$ 
or the reflected wave amplitude; however, in the cases considered below 
there is only one leading quadrature, so different definitions practically
coincide. 

     The corresponding noise-limited sensitivity (minimal detectable charge
for the measurement bandwidth $\Delta f$) is defined as 
$\delta q_0/\sqrt{\Delta f} =\sqrt{S_{Xn}}/|dX_n/dq_0|$ (similarly, 
$\sqrt{S_{Yn}}/|dY_n/dq_0|$), where the low-frequency spectral densities
$S_{Xn}$ and $S_{Yn}$ of quadrature fluctuations are 
        \begin{eqnarray}
&&      S_{Xn}=c_{n}^{\, 2} \langle S_I(t) \sin^2 n\omega t\rangle + 
              d_{n}^{\, 2} \langle S_I(t) \cos^2 n\omega t\rangle 
        \nonumber \\ 
&& \hspace{1cm} - c_{n}d_{n} \langle S_I(t) \sin 2n\omega t\rangle, 
        \label{SXn}\\
&&      S_{Yn}=d_{n}^{\, 2} \langle S_I(t) \sin^2 n\omega t\rangle + 
              c_{n}^{\, 2} \langle S_I(t) \cos^2 n\omega t\rangle 
        \nonumber \\ 
&& \hspace{1cm} +c_{n}d_{n} \langle S_I(t) \sin 2n\omega t\rangle, 
	\label{SYn} \end{eqnarray} 
where 
$c_{n}= (2 R_0 Q n \tilde\omega )/
[n^2\tilde{\omega}^2 + Q^2(1 -n^2\tilde{\omega}^2)^2]$,
$d_{n}= c_n Q (1-n^2 \tilde\omega^2)/n\tilde\omega$, 
$S_I(t)$ is the low-frequency spectral density of the SET shot noise
\cite{Kor-94} with the time dependence due to oscillating bias voltage, 
and the averaging is over the period $2\pi /\omega$.

\begin{figure}
\centerline{ 
\epsfxsize=3.35in
\hspace{-0.2cm}
\epsfbox{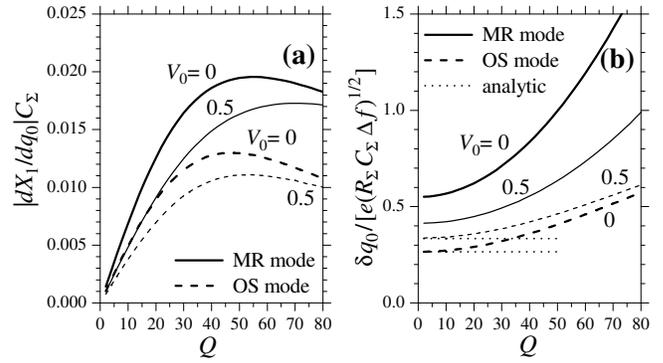} 
}
\vspace{0.2cm} 
\caption{(a) RF-SET response and (b) sensitivity 
as functions of the $Q$-factor in the maximum response (MR) and optimal
sensitivity (OS) modes. $T=0.01\, e^2/C_\Sigma$, $R_\Sigma /R_0=2000$, 
$\omega =\omega_0$. 
}
\label{Qdep}\end{figure}

        Figure \ref{Qdep} shows the numerically calculated RF-SET response
and sensitivity as functions of the ``unloaded'' $Q$-factor 
for a symmetric SET, \cite{asym}
$C_1=C_2=C_\Sigma /2$, $R_1=R_2=R_\Sigma /2$, with $R_\Sigma =100\, 
\mbox{k}\Omega$ at temperature $T=0.01 e^2/C_\Sigma$ for the case of
resonant carrier frequency, $\omega =\omega_0$. Both the response and
sensitivity are shown in respect to the quadrature $X_1$ since all other 
components are small. The RF-SET performance is optimized over the wave 
amplitude $V_{in}$ and the charge $q_0$ to provide either maximum 
response (MR mode; solid lines) or optimized sensitivity (OS mode; dashed
lines). \cite{MR-OS}
 We show the results for two values of the dc bias voltage $V_0$. 
The case $V_0=0$ provides the best MR response and the best OS sensitivity,
and corresponds to the symmetric SET operation in respect to positive and
negative bias voltages (the SET $I-V$ curve is symmetric even for nonzero 
$q_0$). The other shown value $V_0=0.5 e/C_\Sigma$ represents 
a typical case when only one branch of the SET $I-V$ curve is used, 
and corresponds to the plato-like region\cite{Kor-RF} of the response 
and sensitivity dependences on $V_0$. 

   As one can see from Fig.\ \ref{Qdep}(a), the maximum RF-SET response
is achieved 
at $Q$-factors (somewhat different in different regimes) comparable to the
rough estimate  $(R_\Sigma /R_0)^{1/2}\simeq 45$.
However, unlike in the linear model, this maximum does not 
correspond to the exact impedance matching. For example, 
the impedance matching (minimum of reflection) occurs at $Q\simeq 100$ 
for the upper curve in Fig.\ \ref{Qdep}(a) and at $Q\simeq 80$ for 
the curve second from the top, while for two lower curves (OS mode)
it does not occur at all in a reasonable range of $Q$.

 In contrast to the response behavior, the RF-SET sensitivity 
[Fig.\ \ref{Qdep}(b)] monotonically worsens with $Q$.
Qualitatively, this happens because the noise $S_{X1}$ in Eq.\ 
(\ref{SXn}) is proportional to $Q^2$, while the response 
does not grow as fast as $Q$. 
 At low $Q$ the OS sensitivity 
is fitted well by the analytical result \cite{Kor-RF} 
$\delta q_0 \simeq 2.65 e (R_\Sigma C_\Sigma \Delta f)^{1/2} 
(TC_\Sigma /e^2)^{1/2}$ for $V_0=0$ and 
$\delta q_0 \simeq 3.34 C_\Sigma (R_\Sigma T \Delta f)^{1/2}$ 
for the asymmetric operation (shown by dotted lines).
However, at realistic $Q$-factors $\delta q_0$ is significantly larger
(by about 50\% at $Q=50$ for data in Fig.\ \ref{Qdep}). 
Another interesting observation from Fig.\ \ref{Qdep} is that the response 
in the MR mode is only moderately ($\sim 30\%$) better than in the OS mode.

\begin{figure}[t]
\centerline{ 
\epsfxsize=3.06in
\hspace{-0.2cm}
\epsfbox{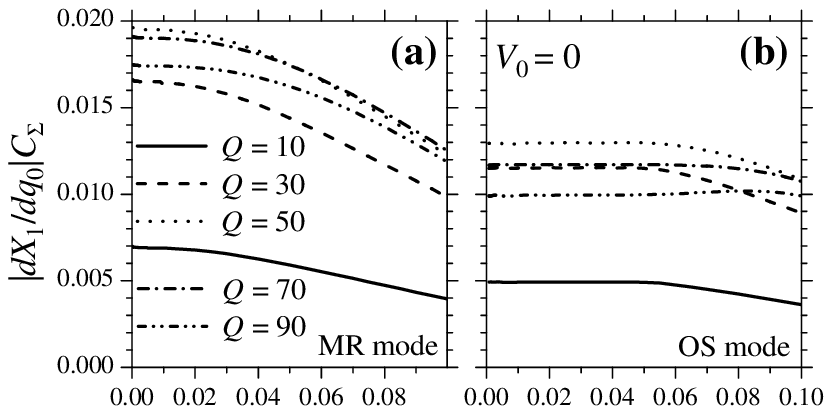} 
}
\vspace{0.1cm}
\centerline{ 
\epsfxsize=3.0in
\epsfbox{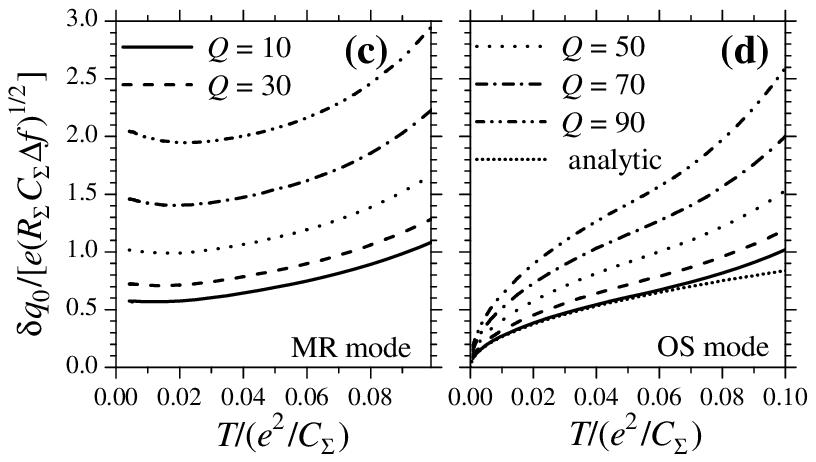} 
}
\vspace{0.2cm}
\caption{Dependence of (a)--(b) RF-SET response and (c)--(d) sensitivity
on temperature $T$ in the MR and OS modes for several $Q$-factors. 
$V_0=0$, $R_\Sigma /R_0=2000$, and $\omega =\omega_0$. 
}
\label{Tdep}\end{figure}

        Figure \ref{Tdep} shows the temperature dependence of the
RF-SET response and sensitivity in the MR and OS modes. 
Even though the low-$T$ analytical formula for the OS sensitivity (above) 
works well only for small $Q$, the $T^{1/2}$ dependence at
$T \lesssim 0.05 e^2/C_\Sigma$ remains valid for large $Q$-factors
(at very small $T$ the OS sensitivity is limited by the neglected here  
contribution from cotunneling processes \cite{Kor-92,Devoret}). 
The OS response practically does not depend on temperature at 
$T \lesssim 0.05 e^2/C_\Sigma$. The performance in the MR mode 
saturates below $T \simeq 0.03 e^2/C_\Sigma$.

\begin{figure}[t]
\centerline{ 
\epsfxsize=3.35in
\hspace{-0.2cm}
\epsfbox{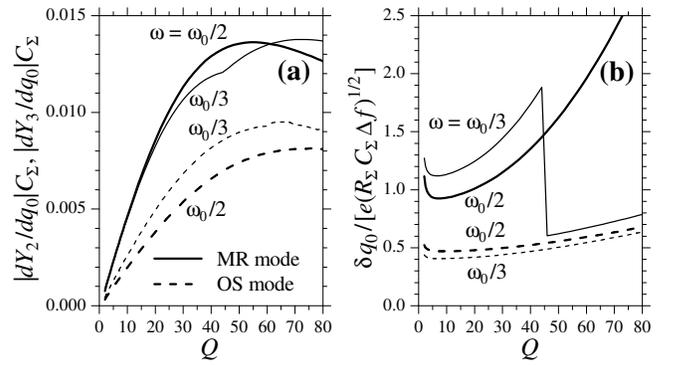} 
}
\vspace{0.2cm}
\caption{(a) RF-SET response and (b) sensitivity in the regimes when the 
second or third overtone of the incident rf wave is in resonance with 
the tank circuit.
$T=0.01\, e^2/C_\Sigma$, $R_\Sigma /R_0=2000$.
}
\label{overtones}
\end{figure}

        So far we have been considering the usual case $\omega =\omega_0$.
In spite of significant SET $I-V$ nonlinearity (the SET nonlinearity 
has been recently used \cite{Knobel} for rf mixing), 
the contribution of overtones in this case is small because they are
off resonance. Even though the formally calculated sensitivities in respect 
to overtones are comparable to the $X_1$ sensitivity (worse by less than 
two times for $n=2$ and 3), the responses are 
much smaller and therefore monitoring of overtones is impractical. 
However, the contribution of $n$th overtone becomes significant if 
$\omega \simeq \omega_0/n$. Figure \ref{overtones} shows the RF-SET 
response and sensitivity for $\omega =\omega_0/2$ 
and $\omega =\omega_0/3$, in respect to monitoring $Y_2$ and $Y_3$, 
correspondingly (the $X$-quadratures are small). We use $V_0=0$
in the case $\omega =\omega_0/3$ and $V_0=0.5 e/C_\Sigma$  in the case 
$\omega =\omega_0/2$ [for $V_0=0$ there is no second overtone because
of the $I-V$ curve symmetry -- see Eq.\ (\ref{anbn})]. 

    Comparing Figs.\ \ref{Qdep} and \ref{overtones} (the parameters
are the same) we see that 
the RF-SET performance in the regime of a resonant 
overtone is comparable to the performance in the conventional regime $\omega 
= \omega_0$ (the MR response and OS sensitivity are worse by about 
1.5 times). 
On the other hand, the frequency separation between the incident wave 
and monitored reflected wave may be an important advantage for 
some applications. 
Also, it may be advantageous 
to have the absence of the monitored wave when the SET is off 
(no current), while for the conventional mode this case corresponds 
to the largest reflected power. 
The disadvantage is a larger incident wave amplitude $V_{in}$ than
for a conventional RF-SET regime, that may lead to heating problems. 
Nevertheless, we hope that the proposed mode of 
the resonant overtone will happen to be practically useful.

\vspace{0.3cm} 
The work was supported by NSA and ARDA under ARO grant 
DAAD19-01-1-0491 and by the SRC grant 2000-NJ-746. 
The numerical calculations were partially performed 
on the UCR-IGPP Beowulf computer Lupin.


\begin{references}
\bibitem{asym} The asymmetric biasing of the RF-SET in the case of 
	significant $C_g$ leads to asymmetric effective capacitances,
	$C_1\neq C_2$. The numerical results show that such asymmetry
	provides a better performance of the RF-SET (that eliminates the 
	concern about asymmetric biasing raised in 
	Ref.\ \protect\cite{Aassime}), though the improvement is quite 
	minor.
\bibitem{MR-OS} Experimentally, the MR mode is preferable if the next stage 
	noise is large; while if it is small, the OS mode is preferable. 
	At low temperatures $V_{in}$ in the OS mode is significantly smaller
 	than in the MR mode.
\end{references}
\end{document}